\documentclass[twocolumn]{aastex62}

\hypersetup{linkcolor=red,citecolor=blue,filecolor=cyan,urlcolor=blue}

\graphicspath{{./}{figures/}}

\shorttitle{The Type I Superluminous Supernova SN\,2016inl}
\shortauthors{Blanchard et al.}

\begin{document}

\title{Late-Time {\it Hubble Space Telescope} Observations of a Hydrogen-Poor Superluminous Supernova Reveal the Power-Law Decline of a Magnetar Central Engine}

\correspondingauthor{Peter K.~Blanchard}
\email{peter.blanchard@northwestern.edu}

\author[0000-0003-0526-2248]{Peter K.~Blanchard}
\affil{Center for Interdisciplinary Exploration and Research in Astrophysics and Department of Physics and Astronomy, Northwestern University, 1800 Sherman Ave.~8th Floor, Evanston, IL 60201, USA}

\author[0000-0002-9392-9681]{Edo Berger}
\affil{Center for Astrophysics \(|\) Harvard \& Smithsonian, 60 Garden St.~Cambridge, MA 02138, USA}

\author[0000-0002-2555-3192]{Matt Nicholl}
\affil{Birmingham Institute for Gravitational Wave Astronomy and School of Physics and Astronomy, University of Birmingham, Birmingham B15 2TT, UK}

\author[0000-0002-7706-5668]{Ryan Chornock}
\affil{Center for Interdisciplinary Exploration and Research in Astrophysics and Department of Physics and Astronomy, Northwestern University, 1800 Sherman Ave.~8th Floor, Evanston, IL 60201, USA}

\author[0000-0001-6395-6702]{Sebastian Gomez}
\affil{Center for Astrophysics \(|\) Harvard \& Smithsonian, 60 Garden St.~Cambridge, MA 02138, USA}

\author[0000-0002-0832-2974]{Griffin Hosseinzadeh}
\affil{Center for Astrophysics \(|\) Harvard \& Smithsonian, 60 Garden St.~Cambridge, MA 02138, USA}

\begin{abstract}
The light curve diversity of hydrogen-poor superluminous supernovae (SLSNe) has kept open the possibility that multiple power sources account for the population.  Specifically, pair-instability explosions (PISNe), which produce large masses of $^{56}$Ni, have been argued as the origin of some slowly-evolving SLSNe.  Here we present detailed observations of SN\,2016inl (=PS16fgt), a slowly-evolving SLSN at $z=0.3057$, whose unusually red spectrum matches PS1-14bj, a SLSN with an exceptionally long rise time consistent with a PISN. Ground-based and {\it Hubble Space Telescope} data, spanning about 800 rest-frame days, reveal a significant light curve flattening, similar to that seen in SN\,2015bn, and much slower than the decline rate expected from radioactive decay of $^{56}$Co. We therefore conclude that despite its slow evolution, SN\,2016inl is inconsistent with a PISN. Instead, the light curve evolution matches the expected power-law spin-down of a magnetar central engine, but with a shallower power law ($L\propto t^{-2.8}$) compared to that in SN\,2015bn, indicating a possible difference in the $\gamma$-ray opacity between the two events.  Analytical modeling indicates typical magnetar engine parameters, but one of the highest ejecta masses ($\approx 20$ M$_{\odot}$) inferred for a SLSN.  Our results indicate that monitoring the late-time light curve evolution of SLSNe provides a powerful diagnostic of their energy source.
\end{abstract}


\section{Introduction} \label{sec:intro}

The power source of hydrogen-poor superluminous supernovae (SLSNe) has remained a topic of intense debate since their discovery \citep{Quimby2011,Chomiuk2011,Gal-Yam2012}.  This is due in part to the diversity of their light curves, most notably a nearly order of magnitude spread in their evolution timescales and peak luminosities \citep{Nicholl2015,DeCia2018,Lunnan2018,Angus2019}.

Radioactive decay of $^{56}$Ni, as in normal hydrogen-poor core-collapse SNe, cannot account for the majority of observed SLSNe given the combination of their high luminosities, light curve timescales, and spectra \citep{Pastorello2010,Inserra2013,Nicholl2014}. Still, a large mass of $^{56}$Ni, possibly produced in a pair-instability supernova \citep[PISN;][]{HegerWoosley2002,Kasen2011}, has been argued as a viable model for the slowest evolving SLSNe (e.g., SN\,2007bi; \citealt{Gal-Yam2009}), although in cases where the rise time has been well measured radioactive decay can be ruled out (e.g., PTF12dam; \citealt{Nicholl2013}).  To date, only a single event with a well measured light curve -- PS1-14bj \citep{Lunnan2016} -- has continued to be a viable candidate for powering by radioactive decay, although even for this event the spectra are inconsistent with the degree of line blanketing expected in PISNe \citep{Kasen2011}. 

Instead, it has been shown that a magnetar engine \citep{KasenBildsten2010,Woosley2010} can account for the light curve diversity of SLSNe \citep{Inserra2013,Nicholl2017}, as well as for their spectra in the photospheric and nebular phases \citep{Dessart2012,Mazzali2016,Nicholl2016,Yan2017b,Jerkstrand2017,Quimby2018,Nicholl2019}. Recently, very late-time optical observations of SN\,2015bn revealed a significant flattening in its light curve to a power-law decline much slower than the radioactive decay of $^{56}$Co, supporting a magnetar central engine \citep{Nicholllate15bn}.

Here, we present observations of a new slowly evolving SLSN, SN\,2016inl at $z=0.3057$, spanning to about 950 observer-frame days post-peak.  SN\,2016inl's early spectra exhibit a close match with the red spectra of PS1-14bj.  Our data clearly reveal a similar light curve flattening as in SN2015bn, with a much slower decline than for $^{56}$Co decay.  Instead, the full light curve can be modeled with a magnetar central engine with parameters typical of the general SLSN population, but a large ejecta mass of $\approx 20$ M$_\odot$.  By association this suggests that PS1-14bj was not a PISN. 

The paper is structured as follows.  In \S\ref{sec:spec} we present the identification of SN\,2016inl as a Type I SLSN with a close match to PS1-14bj, as well as its detailed spectroscopic properties.  In \S\ref{sec:LC} we present and analyze SN\,2016inl's light curve and relevant models.  In \S\ref{sec:host} we present properties of the host galaxy.  In \S\ref{sec:disc} we discuss the nature of the late-time light curve flattening in SN\,2016inl and implications for our understanding of the SLSN population; we conclude in \S\ref{sec:conc}.  Throughout the paper we use $H_{0} = 67$ km s$^{-1}$ Mpc$^{-1}$, $\Omega_{m} = 0.32$, and $\Omega_{\Lambda} = 0.68$ \citep{Planck2013}, resulting in a luminosity distance of 1652 Mpc to SN\,2016inl (for $z=0.3057$ measured from host galaxy emission lines).  The Galactic extinction along the line of sight to SN\,2016inl is $E(B-V) = 0.0658\pm 0.0009$ mag \citep{SF2011}.

\section{A new slowly-evolving SLSN with unusually red spectra}
\label{sec:spec}

\subsection{Discovery and Identification of SN\,2016inl}
SN\,2016inl was detected by the Pan-STARRS Survey for Transients (PSST; \citealt{Huber2015}) on 25 October 2016 (MJD 57686.53) with a magnitude of $19.59\pm 0.03$ in the $w$ filter (dubbed PS16fgt by PSST; \citealt{TNSreport}).  There are no constraining prior non-detections.  We selected it for spectroscopic follow-up because of the lack of a host galaxy at the position of SN\,2016inl in the $grizy$ stacked images of the Pan-STARRS 3$\pi$ Survey \citep{Chambers2016}, leading to a magnitude contrast of $\gtrsim 3.4$ mag.  Our first spectrum from 2016 December 4, obtained with the Blue Channel spectrograph \citep{Schmidt1989} on the 6.5-m MMT telescope, yielded a close match with the near-peak spectrum of PS1-14bj, an unusually red and slowly-evolving SLSN, when shifted to $z\approx 0.30$.  We measure a precise redshift to SN\,2016inl of $z=0.3057\pm 0.0001$ using host galaxy emission lines detected in a later spectrum (\S\ref{specprop}).

\begin{figure*}
\centering
\includegraphics[scale=0.60]{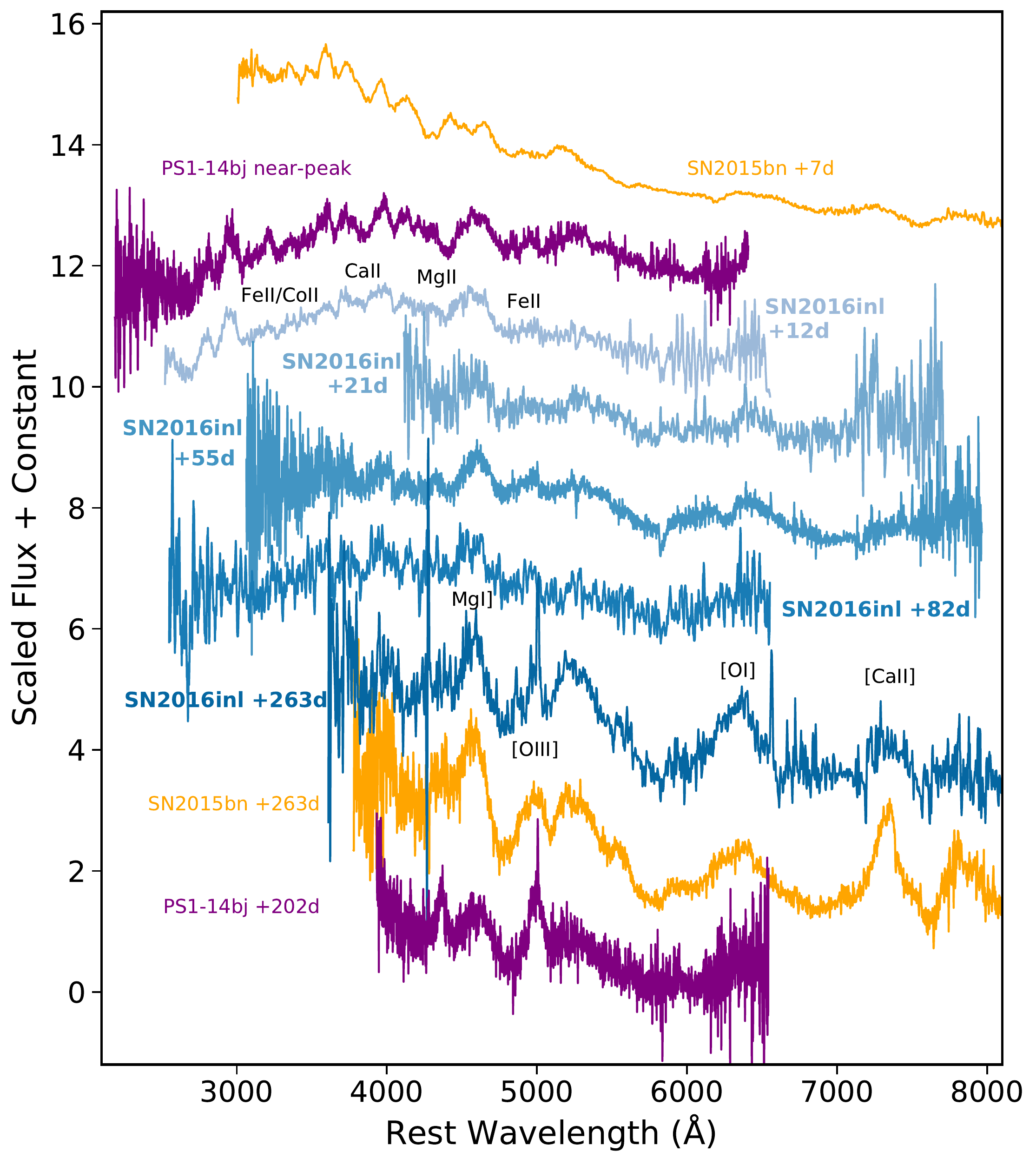}
\caption{Spectra of SN\,2016inl compared with PS1-14bj \citep{Lunnan2016} and SN\,2015bn \citep{Nicholl2016a}.  SN\,2016inl's phases are given relative to the peak time in $w+r$ bands in the rest-frame.  SN\,2016inl and PS1-14bj exhibit a remarkably similar spectrum, with both events being significantly redder than typical SLSNe like SN\,2015bn at a similar phase (see Figure \ref{fig:BB}).  Like PS1-14bj, the spectrum of SN\,2016inl evolves slowly with time, with little change in the spectral shape or features to $+82$ days.  The spectrum at $+263$ days shows emerging nebular emission lines, more similar to SN\,2015bn than PS1-14bj, which showed strong [\ion{O}{3}] lines (see Figure \ref{fig:OIII}).}
\label{fig:spec}
\end{figure*}

\begin{figure}
\centering
\includegraphics[scale=0.36]{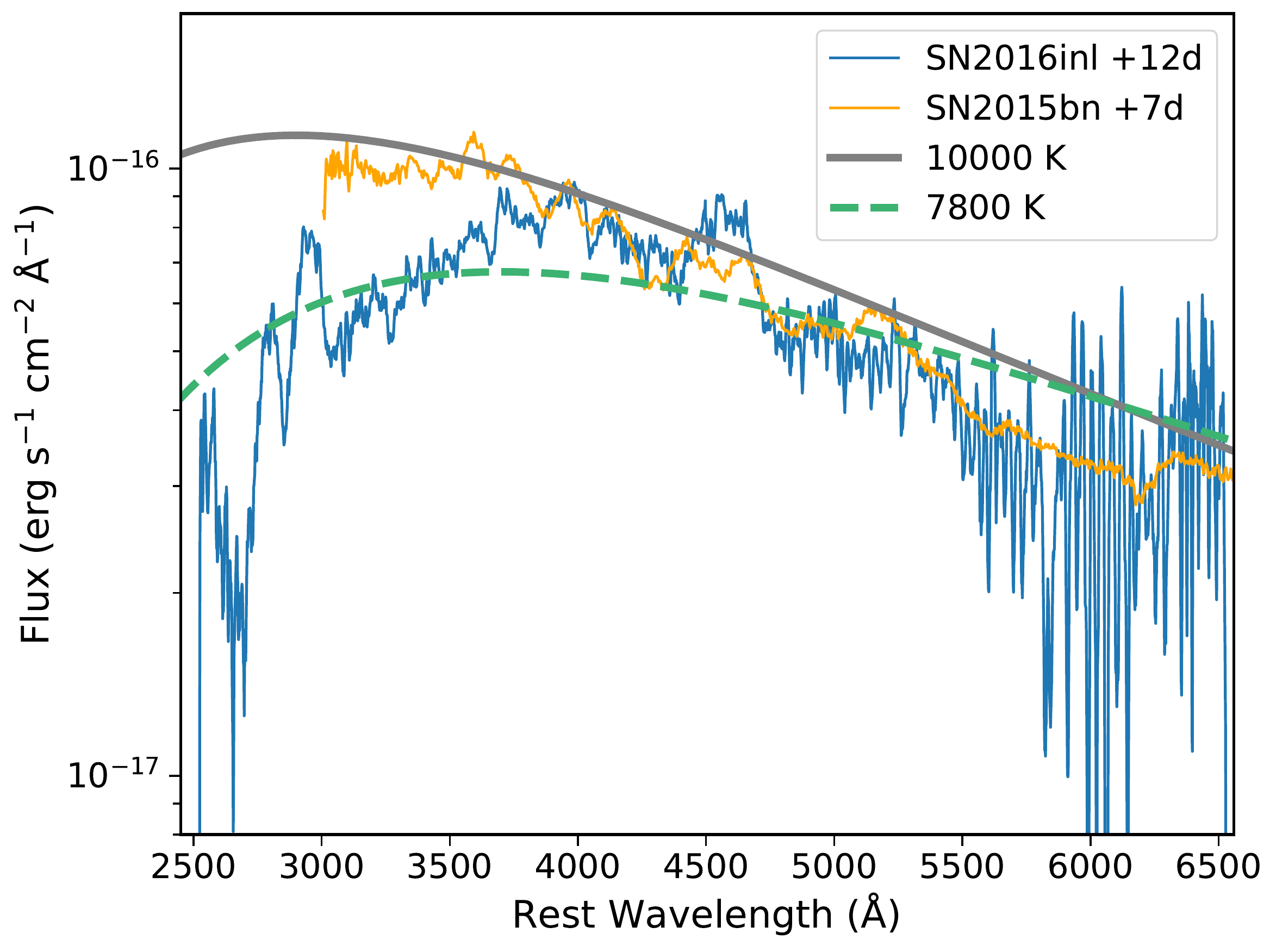}
\caption{Our earliest spectrum of SN\,2016inl compared to a blackbody fit ($T=7800$ K), which is a poor representation of the spectral shape, and a hotter blackbody with $T=10000$ K that matches the spectral shape at $\gtrsim4000$ \AA.  The steep decline blueward of $\approx4000$ \AA\ is therefore likely due to increased line blanketing from Fe-peak elements compared to most other SLSNe (SN\,2015bn is shown).}
\label{fig:BB}
\end{figure}

\begin{figure}
\centering
\includegraphics[scale=0.38]{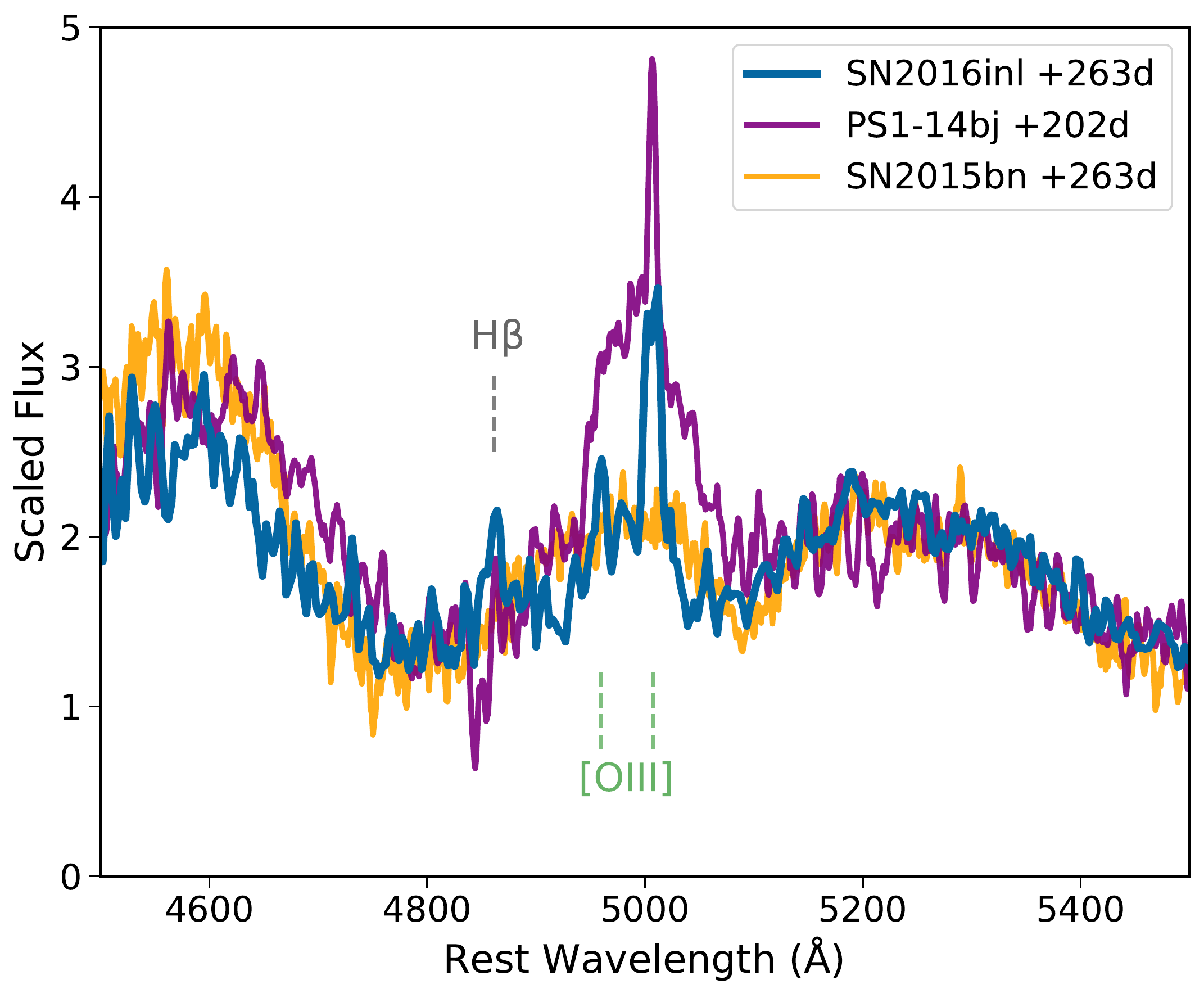}
\caption{Zoom-in on the spectral region containing H$\beta$ and [\ion{O}{3}]$\lambda\lambda$4959,5007 in the $+263$ day spectrum of SN\,2016inl (blue) compared to PS1-14bj (purple) and SN\,2015bn (orange) at similar epochs.  The spectra are scaled to match on the blue and red sides of [\ion{O}{3}].  Narrow emission lines from the host galaxy of SN\,2016inl are detected.  The broad and strong emission line centered at $\approx5000$ \AA\ in PS1-14bj is possibly detected in SN\,2016inl at a lower equivalent width.}   
\label{fig:OIII}
\end{figure}

\begin{figure}
\centering
\includegraphics[scale=0.37]{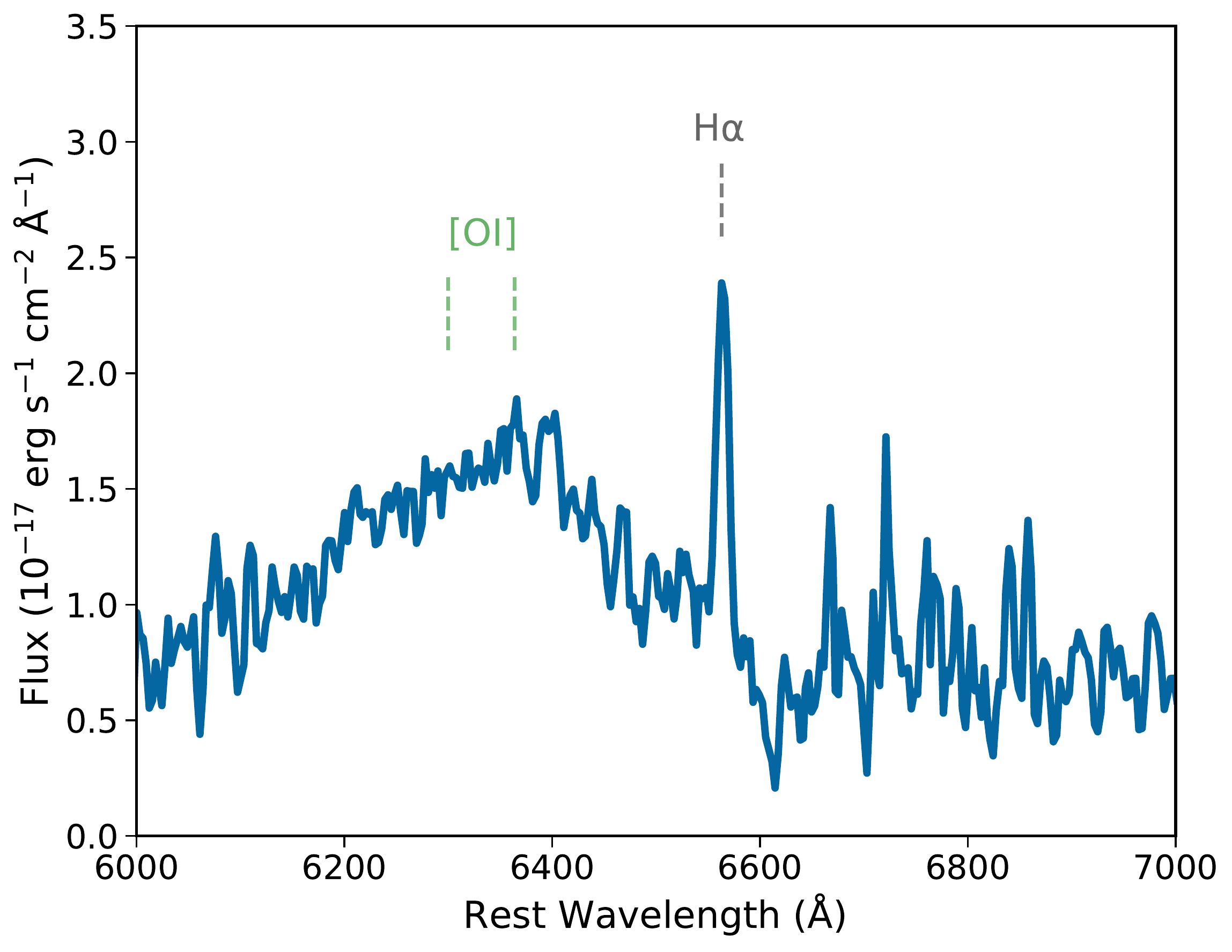}
\caption{Zoom-in on the spectral region containing H$\alpha$ in the $+263$ day spectrum of SN\,2016inl showing that it consists of a single unresolved emission component, originating from the host galaxy. There is no additional broad component in this or any prior spectrum.}
\label{fig:Halpha}
\end{figure}

\subsection{Spectroscopic Observations}

We obtained additional spectroscopic observations of SN\,2016inl using the Blue Channel spectrograph, the Ohio State Multiple Object Spectrograph \citep[OSMOS;][]{Martini2011} on the 2.4-m Hiltner telescope at MDM Observatory, the Low Dispersion Survey Spectrograph \citep[LDSS3c;][]{Stevenson2016} on the 6.5-m Magellan Clay telescope, and the Gemini Multi Object Spectrograph (GMOS; \citealt{Hook2004}) on the Gemini North 8-m telescope.  Details of the observations are presented in Table~\ref{tab:spec}.  We extracted 1D wavelength-calibrated spectra using IRAF and used observations of standard stars obtained on the same nights for relative flux calibration. We used the python code {\tt Pypeit} \citep{Pypeit} to reduce our GMOS observations, which were obtained over two nights in October and November 2017.  To increase the signal-to-noise (S/N) ratio we combined these observations into a single final spectrum.  All times in this section are given in the rest frame relative to MJD 57713.2, our estimate of the peak time in $w+r$ bands (see \S\ref{sec:obsLC}).

\begin{deluxetable*}{ccccccc}[!ht]
\tablecolumns{7}
\tabcolsep0.1in\footnotesize
\tablewidth{7in}
\tablecaption{Spectroscopic Observations of SN\,2016inl  
\label{tab:spec}}
\tablehead {
\colhead {Date (UT)}   &
\colhead {MJD}     &
\colhead {Phase\tablenotemark{a}} &
\colhead {Telescope} &
\colhead{Instrument}  &
\colhead {Airmass}   &
\colhead {Resolution (\AA)}           
}   
\startdata
6 December 2016 & 57728.84 & +12  & MMT & Blue Channel & 1.3 & 4 \\
18 December 2016 & 57740.50 & +21  & MDM/Hiltner & OSMOS & 1.2 & 5 \\
1 February 2017 & 57785.54 & +55  & Magellan/Clay & LDSS3c & 1.9 & 7  \\
8 March 2017 & 57820.62 & +82 & MMT & Blue Channel & 1.6 & 4 \\
31 October 2017\tablenotemark{b} & 58056.50 & +263  & Gemini-North & GMOS-N & 1.1 & 14 \\
\enddata
\tablenotetext{a}{Rest-frame days since MJD 57713.2, our estimate of the peak time in $w+r$ bands.}
\tablenotetext{b}{Two epochs of GMOS observations in October and November 2017 were combined into a single spectrum to increase S/N.  Times here reflect the midpoint time.}
\end{deluxetable*}

\subsection{Spectroscopic Properties}
\label{specprop}

In Figure~\ref{fig:spec} we show the spectral sequence of SN\,2016inl.  SN\,2016inl exhibits a much redder spectrum with lower velocities than typically seen in SLSNe, and moreover shows little evolution during the 70-day coverage of our early spectra.  We identify absorption from intermediate mass elements like \ion{Ca}{2}, \ion{Mg}{2}, and \ion{Fe}{2}, consistent with the phase of our spectra being after peak when the temperature has cooled slightly.  We estimate the photospheric velocity in our $+12$ day spectrum using the \ion{Fe}{2} $\lambda$5169 absorption line, finding the absorption minimum corresponds to $\approx 7000$ km s$^{-1}$.  In addition, we compare this spectrum to a model spectrum calculated using {\tt SYN++} \citep{Thomas2011} and find that a photospheric velocity of 7000 km s$^{-1}$ yields good matches to commonly identified lines from \ion{Ca}{2}, \ion{Mg}{2}, and \ion{Fe}{2}.  This velocity is lower than most previous events which typically have velocities $\gtrsim 10,000$ km s$^{-1}$ \citep{Nicholl2015,Liu2017,Lunnan2018,Quimby2018}. 

In Figure \ref{fig:BB} we show a blackbody fit to the spectrum at 12 days, which has a best-fit temperature of $\approx7800$ K.  However, the overall shape of the spectrum is not well-approximated by a single blackbody due to the steep drop-off in flux blueward of $\approx 4000$ \AA.  The spectral shape at $\gtrsim4000$ \AA, on the other hand, is well-approximated by a $T=10000$ K blackbody, similar to the temperatures of other SLSNe more than a week after peak \citep{Inserra2013,Nicholl2015,Lunnan2018}.  For comparison we show the spectrum of SN\,2015bn at 7 days after peak \citep{Nicholl2016a}, which, like most SLSNe, does not show the same level of suppression in the blue.  While SLSNe are known to exhibit significant absorption in the UV, in most events the absorption is only significant blueward of $\sim 3000$ \AA\ \citep{Chomiuk2011,Vreeswijk2014,Yan2017a,Nicholl2017a,Quimby2018}.  Therefore the red spectrum of SN\,2016inl is likely due to increased line blanketing rather than a cooler underlying continuum.  SN\,2016inl appears remarkably similar to PS1-14bj \citep{Lunnan2016}, a SLSN with a broad light curve, which also shows a steep decline in flux blueward of $\approx 4000$ \AA.  Evidence of increased absorption in the blue has also been seen in SN\,2017dwh \citep{Blanchard2019}.  As demonstrated in the {\tt SYN++} models of \citet{Gal-Yam2009}, \citet{Lunnan2016}, and \citet{Blanchard2019}, and others in the context of normal SNe (e.g., \citealt{Mazzali2000,Mazzali2002,Mazzali2017,Sasdelli2014}), absorption from ionized Fe and Co can explain flux suppression in the region $\approx2800-3500$ \AA.  Models presented by \citet{Quimby2018} show that the addition of \ion{Fe}{2} leads to a strong line near 3050 \AA, which is present in both PS1-14bj and SN\,2016inl.  Increased absorption from Fe-peak elements is therefore likely a significant factor explaining the flux suppression at wavelengths below $\approx4000$ \AA\ in SN\,2016inl.      

The spectrum of SN\,2016inl at 263 rest-frame days after peak still shows a residual continuum, while the developing nebular emission lines indicate the transition to the nebular phase.  Aside from a weaker [\ion{Ca}{2}]$\lambda$7300 doublet, SN\,2016inl at this phase is similar to SN\,2015bn.  SLSNe generally do not become fully nebular until $\gtrsim 1$ year after peak, with some variation associated with the diversity in light curve evolution timescales \citep{Nicholl2019}.  In the 263 day spectrum we also detect host galaxy emission lines (H$\alpha$, H$\beta$, and [\ion{O}{3}]) from which we determine the redshift.  

The spectrum at 263 days also exhibits a broad emission line near 5000 \AA, which was seen at greater strength in PS1-14bj starting as early as 155 rest-frame days after peak and attributed to [\ion{O}{3}] (Figure~\ref{fig:OIII}).  We fit a model consisting of three Gaussian profiles, two fixed at the rest-frame locations of the [\ion{O}{3}] doublet lines with widths fixed at the resolution of the spectrograph to account for the host emission lines.  The best-fit third component is centered at 4989 \AA\ with a full-width at half maximum (FWHM) of $\approx 5000$ km s$^{-1}$, consistent with the FWHM of the profile seen in PS1-14bj.  However, given that SN\,2016inl is not completely in the nebular phase at 263 days this broad spectral peak underlying the narrow host lines may be the result of nearby absorption from \ion{Fe}{2}.  As most SLSNe do not show broad [\ion{O}{3}] emission lines as strong as those seen in PS1-14bj \citep{Nicholl2019}, at this phase SN\,2016inl is more consistent with the general population than PS1-14bj.  

Finally, we find no spectral signatures of circumstellar interaction at any phase.  In particular, there is no hydrogen emission evident at any phase, other than that associated with the host galaxy.  As some SLSNe have shown emerging broad H$\alpha$ emission at late times \citep{Yan2017b}, we carefully examine our late-time GMOS spectrum.  In Figure \ref{fig:Halpha}, we show a zoom-in around H$\alpha$ in the $+263$ day spectrum where we find no indication of broad emission underlying the unresolved host H$\alpha$ line.  

\begin{figure*}[th]
\centering
\includegraphics[scale=0.6]{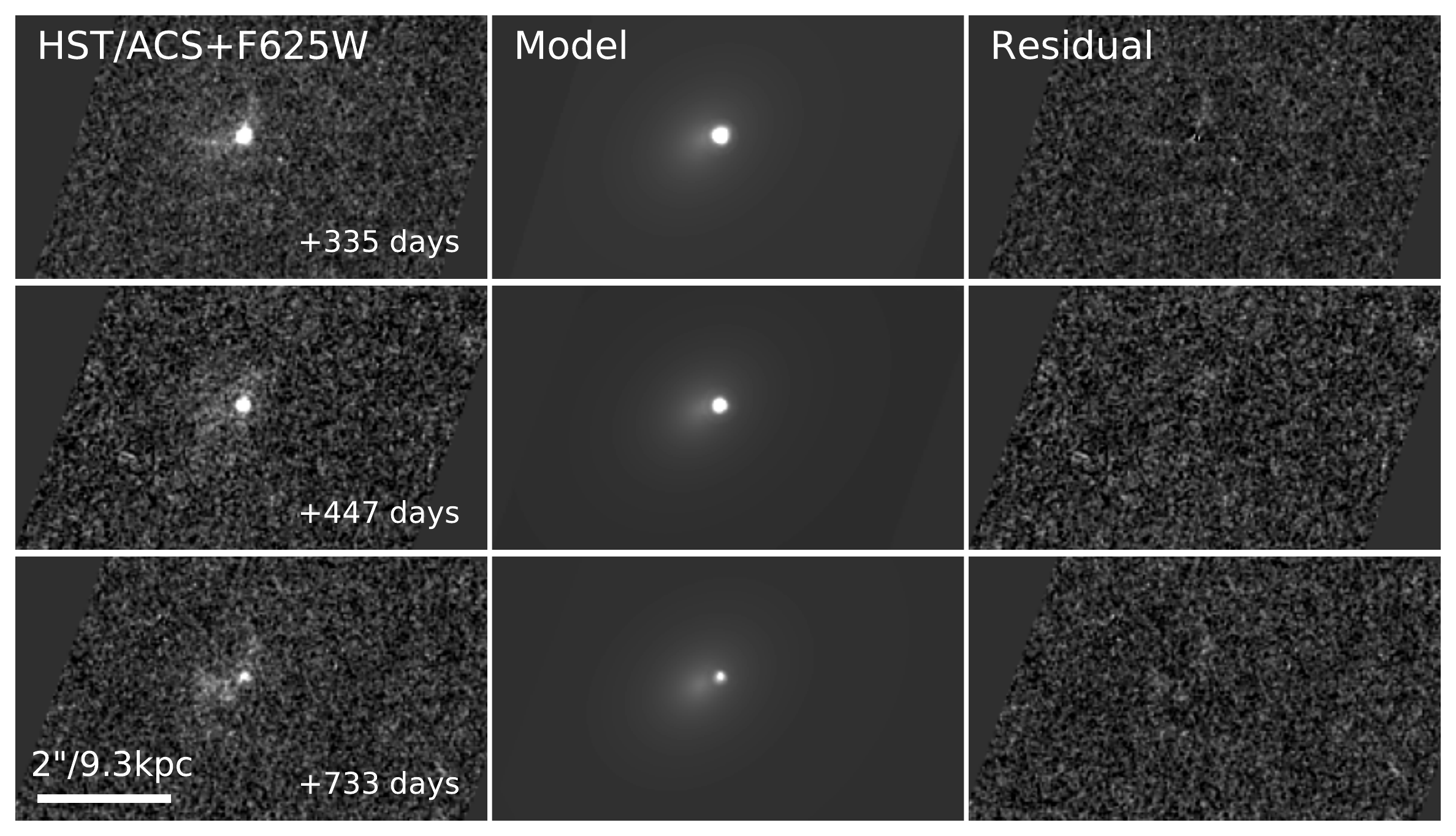}
\caption{\textit{HST}/ACS observations of SN\,2016inl at 335, 447, and 733 rest-frame days after $r$-band peak ({\it Left}) with corresponding GALFIT model ({\it Middle}) and residual ({\it Right}) images.  By simultaneously fitting the image with a Sersic profile (galaxy) and PSF (SN), we are able to robustly separate the SN flux from underlying host emission.
\label{fig:HST}}
\end{figure*}

\begin{deluxetable}{cccc}
\tablecolumns{4}
\tabcolsep0.05in\footnotesize
\tablewidth{4in}
\tablecaption{Photometry of SN\,2016inl
\label{tab:phot}}
\tablehead {
\colhead {MJD}   &
\colhead{Magnitude\tablenotemark{a}} &
\colhead{Filter} &
\colhead{Instrument}
}   
\startdata
57686.53\tablenotemark{b} & 19.41 $\pm$ 0.02 & $w$ & PS1/GPC1 \\
57726.18 & 19.31 $\pm$ 0.10 & $r$ & FLWO/KeplerCam \\
57737.18 & 19.62 $\pm$ 0.22 & $g$ & FLWO/KeplerCam \\
57746.35 & 19.82 $\pm$ 0.20 & $g$ & FLWO/KeplerCam \\
57746.35 & 19.42 $\pm$ 0.18 & $i$ & FLWO/KeplerCam \\
57749.35 & 19.78 $\pm$ 0.18 & $g$ & FLWO/KeplerCam \\
57749.35 & 19.48 $\pm$ 0.10 & $r$ & FLWO/KeplerCam \\
57749.35 & 19.56 $\pm$ 0.11 & $i$ & FLWO/KeplerCam \\
57759.12 & 20.12 $\pm$ 0.18 & $g$ & FLWO/KeplerCam \\
57759.12 & 19.58 $\pm$ 0.10 & $r$ & FLWO/KeplerCam \\
57759.12 & 19.56 $\pm$ 0.12 & $i$ & FLWO/KeplerCam \\
57771.27 & 20.22 $\pm$ 0.15 & $g$ & FLWO/KeplerCam \\
57771.27 & 19.76 $\pm$ 0.09 & $r$ & FLWO/KeplerCam \\
57771.27 & 19.75 $\pm$ 0.11 & $i$ & FLWO/KeplerCam \\
57785.54 & 20.06 $\pm$ 0.05 & $r$ & Magellan/LDSS3c \\
57807.10 & 20.33 $\pm$ 0.08 & $r$ & FLWO/KeplerCam \\
57956.89 & 22.44 $\pm$ 0.08 & $r$ & Magellan/LDSS3c \\
57956.89 & 22.76 $\pm$ 0.07 & $g$ & Magellan/LDSS3c \\
57956.89 & 22.73 $\pm$ 0.14 & $i$ & Magellan/LDSS3c \\
57956.89 & $>21.93$ & $z$ & Magellan/LDSS3c \\
58014.31 & 23.20 $\pm$ 0.33 & $g$ & MMT/MMTCam\\
58014.31 & 22.48 $\pm$ 0.20 & $r$ & MMT/MMTCam\\
58046.79 & 23.25 $\pm$ 0.10 & $g$ & Magellan/IMACS \\
58046.79 & 22.66 $\pm$ 0.10 & $r$ & Magellan/IMACS \\
58046.79 & 23.10 $\pm$ 0.13 & $i$ & Magellan/IMACS \\
58046.79 & 22.65 $\pm$ 0.22 & $z$ & Magellan/IMACS \\
58073.38 & 23.46 $\pm$ 0.12 & $r$ & MMT/Binospec \\
58073.38 & 23.83 $\pm$ 0.12 & $g$ & MMT/Binospec \\
58073.38 & $>21.93$ & $z$ & MMT/Binospec \\
58111.56 & 23.88 $\pm$ 0.19 & $g$ & Magellan/IMACS \\
58111.56 & 23.54 $\pm$ 0.15 & $r$ & Magellan/IMACS \\
58111.56 & 23.32 $\pm$ 0.21 & $i$ & Magellan/IMACS \\
58151.12 & 24.11 $\pm$ 0.05 & F625W & {\it HST}/ACS \\
58297.03 & 24.50 $\pm$ 0.05 & F625W & {\it HST}/ACS \\
58430.14 & $>25.02$ & $r$ & MMT/Binospec \\
58670.86 & 26.09 $\pm$ 0.16 & F625W & {\it HST}/ACS  
\enddata
\tablenotetext{a}{AB and corrected for Galactic extinction.}
\tablenotetext{b}{Reported by PSST \citep{TNSreport}.}
\end{deluxetable}

\section{Light Curve and Modeling} 
\label{sec:LC}

\subsection{Imaging Data}
We obtained images of SN\,2016inl in the $griz$ filters using KeplerCam on the 48-inch telescope at Fred Lawrence Whipple Observatory (FLWO), the Inamori-Magellan Areal Camera and Spectrograph \citep[IMACS;][]{Dressler2011} on the 6.5-m Magellan Baade telescope, LDSS3c on the 6.5-m Magellan Clay telescope, and MMTCam and Binospec \citep{Fabricant2019} on the 6.5-m MMT telescope.  These images were analyzed using standard techniques in IRAF.  

We also obtained three epochs of imaging with the Advanced Camera for Surveys (ACS) on the {\it Hubble Space Telescope} ({\it HST}) in the F625W filter (PI: Blanchard, Program IDs: 15162 and 15496).  Each {\it HST} epoch consisted of one orbit of four dithered ACS images, which were drizzle-combined using the {\tt astrodrizzle} task ({\tt final\_pixfrac} = 0.8 and {\tt final\_scale} = 0.035''/pixel) in the DrizzlePac\footnote{\url{https://www.stsci.edu/scientific-community/software/drizzlepac.html}} software package.    

Photometry on the ground-based images obtained during the first observing season ($\lesssim$ 100 observer-frame days post peak) was performed using PSF fitting implemented by the IRAF {\tt daophot} package.  Instrumental magnitudes were calibrated to AB magnitudes on the PS1/3$\pi$ photometric system using comparison stars in the field.  In these images, the contribution from the host galaxy is negligible and so host subtraction was not necessary. 

Photometry on images obtained during the second observing season ($\gtrsim$ 200 observer-frame days after peak), when the SN had faded by several magnitudes, was performed after careful consideration of the host galaxy contribution.  Late-time ground-based observations yielded a slightly extended source, indicating a faint, compact host galaxy underlying the SN.  In our deep, high-resolution {\it HST} observations the emission at the position of SN\,2016inl is clearly resolved into both a point source and a diffuse host galaxy component (Figure~\ref{fig:HST}).  For the purpose of image subtraction with the ground-based data (using {\tt HOTPANTS}; \citealt{Becker2015}), we obtained $griz$ templates in December 2019 and January 2020 ($\sim$1140 days after peak).  Image subtraction performed on data from November 2018 ($\sim$720 days after peak) showed no residual flux, indicating no significant SN flux in deep ground-based data by this time.  

To measure the SN flux in the {\it HST} images, we simultaneously model the flux from SN\,2016inl and its host galaxy with PSF and Sersic profiles, respectively, using GALFIT \citep{Peng2010}.  We show the model and residual images in Figure~\ref{fig:HST}, demonstrating that this procedure provides a reliable way to disentangle the flux from the SN and its host galaxy.  We calibrate the {\it HST} photometry using the published ACS/F625W zeropoint.

\subsection{Light Curve Behavior}
\label{sec:obsLC}

\begin{figure*}[t]
\centering
\includegraphics[scale=0.55]{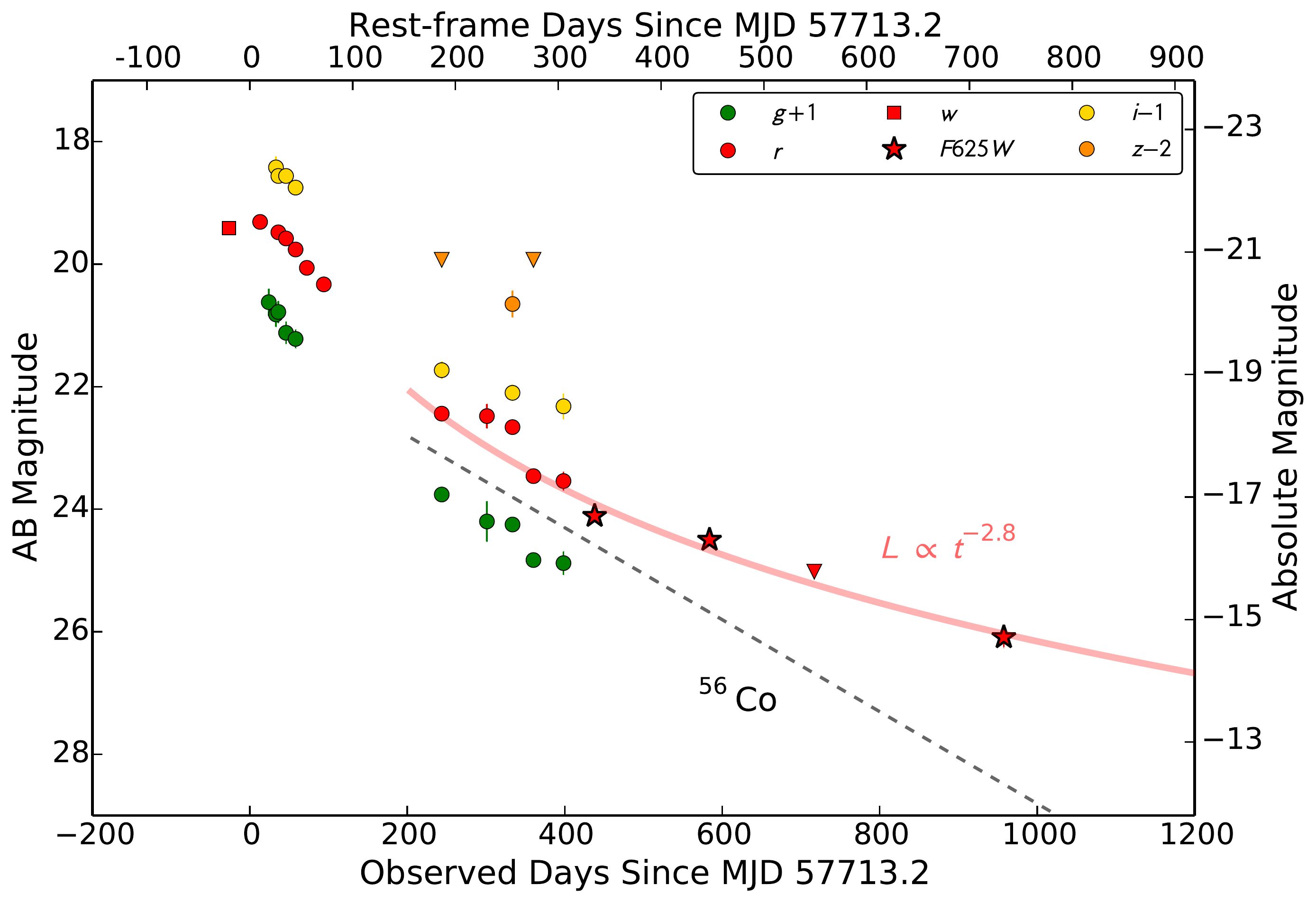}
\caption{Multi-band light curves of SN\,2016inl (relative to the peak in $w+r$; corrected for Galactic extinction) including the PSST $w$-band observation, our ground-based $griz$ observations, and our {\it HST} ACS/WFC F625W observations. Also shown is the decline rate expected from fully-trapped radioactive decay of $^{56}$Co (black dashed line).  The observed late-time evolution of the light curve (especially in the {\it HST} data) is clearly shallower than expected for $^{56}$Co decay.  The flattening behavior is well-represented by a power law; a fit to the combined $r$+F625W light curve at $\gtrsim200$ days is shown as a red line.}
\label{fig:obsLC}
\end{figure*}

The light curves of SN\,2016inl are shown in Figure~\ref{fig:obsLC} and the photometry is listed in Table~\ref{tab:phot}.  By the time we began follow-up observations, SN\,2016inl was already declining in brightness. However, the discovery $w$-band observation from PSST provides a constraint on the time of peak in $r$-band, since it is $\approx 0.5$ mag dimmer than an extrapolation of our early $r$-band data and the zeropoints of $r$ and $w$ do not differ significantly.  We estimate the time of peak of the combined $w+r$ light curve by fitting a polynomial to the data at $\lesssim70$ days, which gives MJD 57713.2.  The times in this section are given relative to this time.  SN\,2016inl maintained a consistent decline rate in $r$-band of $0.018 \pm 0.001$ mag d$^{-1}$ in the rest frame until becoming Sun constrained.  When observations resumed the decline rate decreased substantially, averaging $\approx0.012$ mag d$^{-1}$ at $\approx 180-330$ rest-frame days after peak with evidence for relatively rapid variability at $\approx 260$ days.  At $\gtrsim 330$ rest-frame days after peak, the decline rate slowed further to $0.0050\pm 0.0004$ mag d$^{-1}$ as determined from a fit to the \textit{HST}/F625W data, which span $\approx 330-730$ rest-frame days after peak.

In Figure~\ref{fig:obsLC} we show the exponential decline rate of $\approx$ 0.0098 mag d$^{-1}$ for fully-trapped radioactive decay of $^{56}$Co.  The light curve decline of SN\,2016inl slows considerably relative to $^{56}$Co decay and instead exhibits a decline reminiscent of a power law.  We fit the power-law function $L \propto t^{-\alpha}$ to the combined $r$+F625W light curve at $\gtrsim200$ days after peak and find a best-fit power-law index of $\alpha=2.8\pm0.2$.  The best fit is shown in Figure \ref{fig:obsLC}.  Aside from the relatively rapid drop in flux at $\approx260$ days, the fit adequately captures the overall behavior spanning more than 500 days, and particularly the data at $\gtrsim300$ days.

\subsection{Light Curve Modeling}

\subsubsection{Magnetar Model}
\label{modeling}

\begin{figure*}
\centering
\includegraphics[scale=0.55]{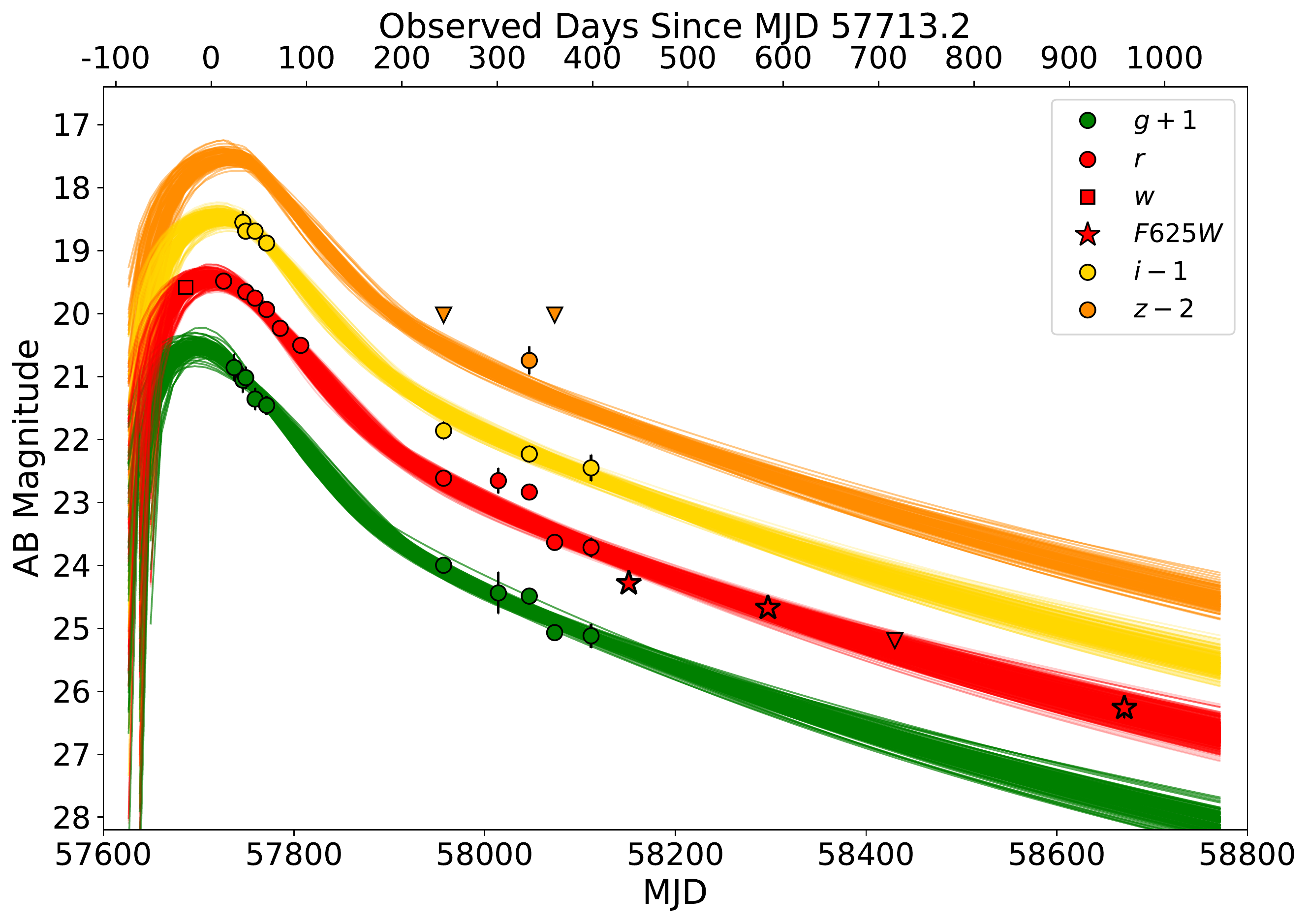}
\caption{Light curves of SN\,2016inl compared to an ensemble of magnetar model realizations calculated using {\tt MOSFiT}.  Unlike radioactive decay models, the late-time flattening is easily accommodated by the power-law spin-down of a magnetar engine with parameters similar to those of other SLSNe.}
\label{fig:mosfit}
\end{figure*}

\begin{deluxetable}{cc}[t!]
\tablecolumns{2}
\tabcolsep0.32in\footnotesize
\tablewidth{4in}
\tablecaption{Magnetar Engine Model Parameters  
\label{tab:param}}
\tablehead {
\colhead {Parameter}   &
\colhead {Value$^a$} 
}   
\startdata
$P_{\rm spin}$ (ms) & 1.48$^{+0.38}_{-0.28}$  \\[5pt]
$B$ (10$^{14}$ G) & $2.15^{+3.39}_{-1.22}$  \\[5pt]
$M_{\rm ej}$ (M$_{\odot}$) & 19.95$^{+6.96}_{-4.82}$  \\[5pt] 
$v_{\rm ej}$ (km s$^{-1}$) & 6900$^{+690}_{-630}$  \\[5pt] 
$E_{\rm k}$ (10$^{51}$ erg) & 8.90$^{+5.18}_{-3.07}$  \\[5pt]
$t_{\rm exp}$ (days) & $-56.2^{+5.8}_{-8.5}$  \\[5pt] 
$\kappa$ (cm$^{2}$ g$^{-1}$) & 0.18$^{+0.01}_{-0.02}$  \\[5pt]
log $\!\kappa_{\gamma}$ & $-0.94^{+0.08}_{-0.04}$  \\[5pt]
$M_{\rm NS}$ (M$_{\odot}$) & 1.63$^{+0.28}_{-0.36}$  \\[5pt]
$T_{\rm min}$ (K) & 8670$^{+440}_{-380}$  \\[5pt]
log($A_{\rm V}^{\rm host}$) & $-2.59^{+1.60}_{-1.55}$  \\[5pt]
log $\!\sigma$ & $-0.94^{+0.08}_{-0.07}$ 
\enddata
\tablenotetext{a}{Median and $\pm1\sigma$ ranges corresponding to the model posteriors.}
\tablecomments{$P_{\rm spin}$ is the initial spin period of the magnetar, $B$ is the component of the magnetar magnetic field perpendicular to the spin axis, $M_{\rm ej}$ is the ejecta mass, $v_{\rm ej}$ is the ejecta velocity, $E_{\rm k}$ is the kinetic energy, $\kappa$ is the opacity, $\!\kappa_{\gamma}$ is the gamma-ray opacity, $M_{\rm NS}$ is the neutron star mass, $T_{\rm min}$ is the photospheric temperature floor, $A_{\rm V}^{\rm host}$ is the internal host galaxy extinction, $t_{\rm exp}$ is the explosion time relative to the first observation, and $\!\sigma$ is the uncertainty required to yield a reduced chi-squared of 1.  For more details on the model and these parameters see \citet{Nicholl2017}.} 
\end{deluxetable}

\begin{figure*}
    \centering
    \includegraphics[scale=0.45]{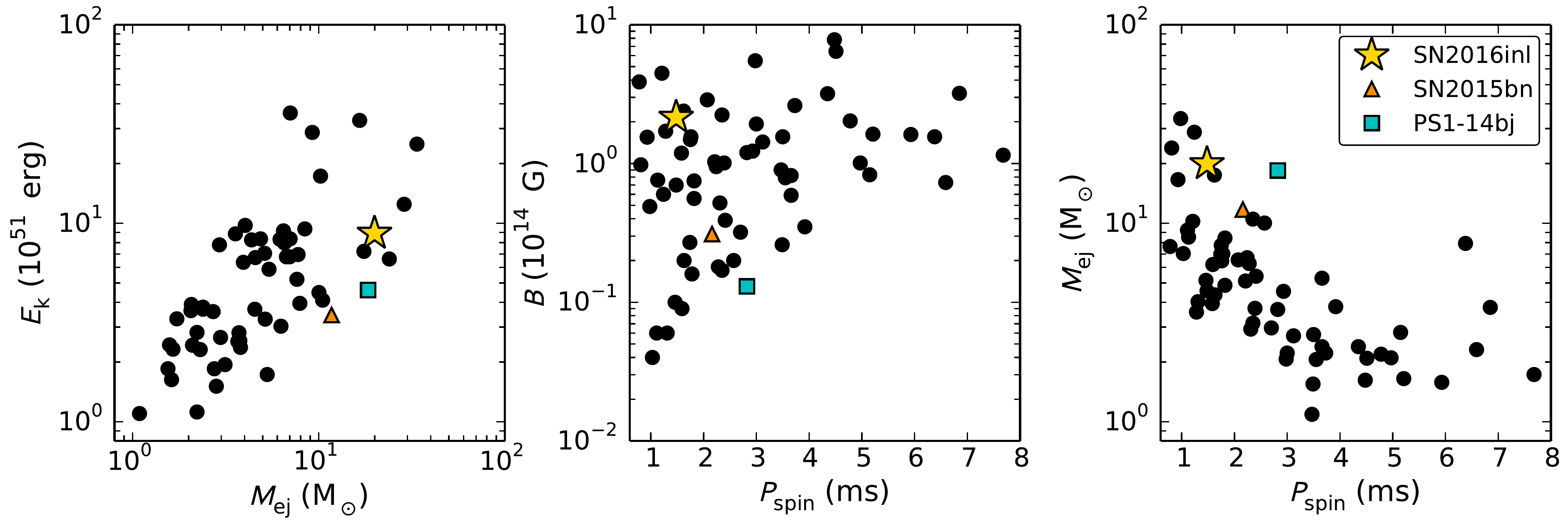}
    \caption{Comparison of the engine and ejecta parameters for SN\,2016inl (yellow star) to the large sample of SLSNe modelled with {\tt MOSFiT} (black; \citealt{Nicholl2017,Villar2018,Blanchard2018,Blanchard2019,Blanchard2020}).  We also mark specifically the relevant comparison objects PS1-14bj and SN\,2015bn.}
    \label{fig:param}
\end{figure*}

We use the Modular Open Source Fitter for Transients \citep[{\tt MOSFiT};][]{Guillochon2018}, a Markov Chain Monte Carlo (MCMC) light curve fitting code, to model SN\,2016inl with the magnetar engine model, where the energy input is solely provided by the spin-down of a central magnetar.  We use the same model and priors as in \citet{Nicholl2017}. We ran the MCMC fitting procedure until convergence, as assessed using the condition that the Potential Scale Reduction Factor is $<1.1$ \citep{GelmanRubin1992,BrooksGelman1998}.  This typically equates to about $25,000-40,000$ iterations.  The magnetar model realizations are shown in Figure~\ref{fig:mosfit}, and provide excellent fits to the data, accounting for the evolving shape of the light curve.  The medians of the parameter posterior distributions and associated $1\sigma$ uncertainties are listed in Table~\ref{tab:param}.  We find a large ejecta mass of $M_{\rm ej}\approx 20$ M$_{\odot}$, consistent with the broad light curve, and a velocity of $v_{\rm ej}\approx 6900$ km s$^{-1}$, consistent with the spectra.  The kinetic energy is large, $E_{\rm k}\approx 9\times 10^{51}$ erg.  The engine parameters, $P_{\rm spin}\approx 1.5$ ms and $B\approx 2\times 10^{14}$ G, are typical for SLSNe, as shown in Figure~\ref{fig:param}.  We therefore conclude that SN\,2016inl is mainly distinguished from most SLSNe by a large ejecta mass (Figure~\ref{fig:param}).  We note that SN\,2016inl exists in a different part of engine parameter space than PS1-14bj; we discuss the implications of this in \S\ref{sec:disc}.

We also find that the explosion date is about 56 observer-frame days prior to the first detection, which corresponds to a rise time of $\approx 64$ rest-frame days to our estimate of the peak time in $w+r$ (\S\ref{sec:obsLC}).  However, due to the high temperatures and rapid temperature evolution predicted by the models at early times, bolometric maximum is predicted to have occurred before the peak in $r$-band, at $\approx37$ rest-frame days after explosion.  As discussed in \S\ref{sec:disc}, this is key for understanding the spectroscopic evolution in relation to the engine properties of SN\,2016inl and our comparison objects.  

\subsubsection{Radioactive Decay Model}

While we have shown the late-time decline of SN\,2016inl follows a power law, significantly slower than the exponential decay of $^{56}$Co, here we fit radioactive decay models for a robust comparison.  Pair-instability supernova models, which produce $\gtrsim$\,10 M$_{\odot}$ of $^{56}$Ni \citep{Kasen2011}, predict long evolution timescales ($\gtrsim 100$ day rise times) due to the large ejecta masses, and have therefore been considered as a possible explanation for some slowly-evolving SLSNe.  In cases like PS1-14bj and SN\,2016inl with either long or poorly constrained rise times, late-time data are critical for assessing the viability of a PISN interpretation.  Observations of PS1-14bj extend to only $\approx 400$ rest-frame days after peak, leaving open the possibility of a PISN interpretation \citep{Lunnan2016}, though a magnetar was favored due to a poor match with spectral models of PISNe.

We fit the multi-band light curves of SN\,2016inl with a radioactive-decay powered model using {\tt MOSFiT}.  We parameterize the $^{56}$Ni mass as a fraction, $f_{\rm Ni}$, of the total ejecta mass, and use a flat prior of $f_{\rm Ni} = 0 - 1$. As expected, we find poor fits to the data, which cannot account for both the post-peak decline rate and the late-time flattening.  We find an ejecta mass of $\gtrsim 30$ M$_{\odot}$ and $^{56}$Ni mass of $\gtrsim 20$ M$_{\odot}$ with precise values depending on the prior on the uncertain explosion epoch.  Such a large nickel fraction of $\gtrsim 50$\% is atypical for core-collapse SNe.  Allowing for long rise times, we find fits with large ejecta masses of $\gtrsim 100$ M$_{\odot}$, effectively producing PISN models similar to those in \citet{Kasen2011}.  

In radioactive models, the late-time decline is limited by the decay of $^{56}$Co until the decay of isotopes with longer half-lives, such as $^{57}$Co, start to dominate the luminosity.  However, for reasonable ratios of $^{57}$Co/$^{56}$Co (e.g., 0.058 as in SN\,1987A; \citealt{Seitenzahl2014}), this transition does not occur until $\gtrsim 1000$ days after explosion, whereas the flattening in SN\,2016inl begins at $\sim 350$ days.  Explaining the late-time luminosity of SN\,2016inl with the decay of $^{57}$Co would also require an unrealistically large mass of $\gtrsim 10$ M$_{\odot}$, a factor of $\gtrsim 10^3$ times higher than in SN\,1987A \citep{Seitenzahl2014}.  Even in the most massive PISN models that produce $\gtrsim 40$ M$_{\odot}$ of $^{56}$Ni, $\lesssim1$ M$_{\odot}$ of $^{57}$Ni, the parent isotope of $^{57}$Co, is predicted \citep{HegerWoosley2002}.   

We therefore conclude that radioactive decay alone cannot explain the light curve of SN\,2016inl.

\subsection{Bolometric Light Curve}

\begin{figure*}
\centering
\includegraphics[scale=0.45]{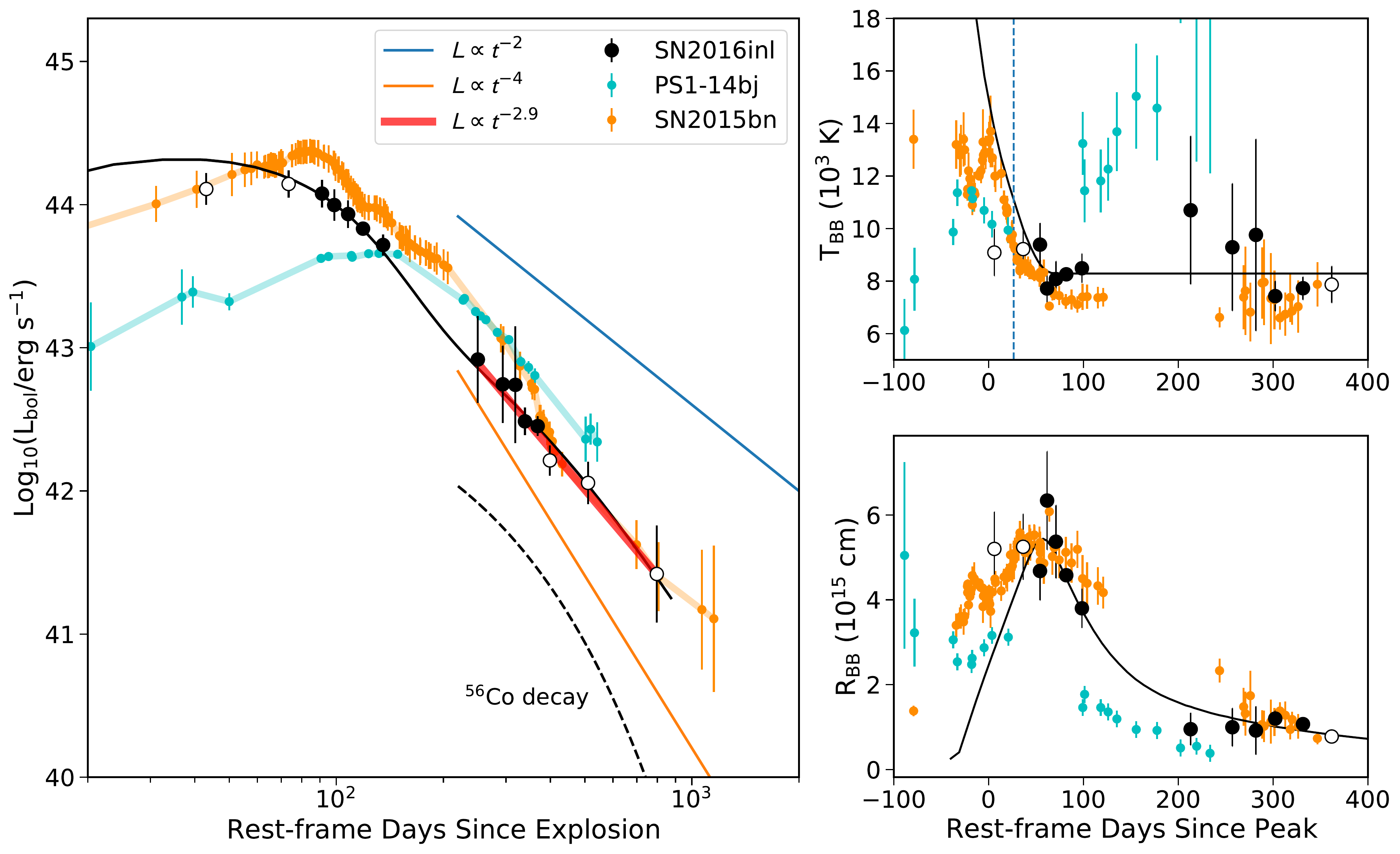}
\caption{{\it Left:} Bolometric light curve of SN\,2016inl compared to those of PS1-14bj \citep{Lunnan2016} and SN\,2015bn \citep{Nicholllate15bn}.  SN\,2016inl exhibits a similar power-law decline as SN\,2015bn, though with a shallower slope (best-fit $L \propto t^{-2.9}$; red line), making it only the second SLSN to have sufficient data to show a decline rate substantially slower than $^{56}$Co decay (black dashed line).  We also show magnetar spin-down luminosity ($L \propto t^{-2}$; blue line) and magnetar spin-down luminosity with high-energy leakage assuming a constant $\gamma$-ray opacity ($L\propto t^{-4}$; orange line).  {\it Top Right:} Photospheric temperature evolution of SN\,2016inl inferred from blackbody fits to the $gri$ light curves.  The vertical line marks the time of peak in $r$-band.  {\it Bottom Right:} Corresponding evolution of the photospheric radius.  The right panels are shown relative to the time of bolometric maximum, which for SN\,2016inl we infer from our magnetar modeling.  We also show the bolometric light curve and corresponding temperature and radius evolution of the best-fit magnetar model from \S\ref{modeling} (solid black lines).  The open symbols indicate epochs with single-band data in which we extrapolated the other bands assuming constant colors relative to the nearest epoch with multi-band data.
\label{fig:bolLC}}
\end{figure*}

We use the python code {\tt SuperBol} \citep{superbol} to calculate the rest-frame bolometric light curve of SN\,2016inl by integrating the observed flux and fitting a blackbody to the $gri$ light curves to estimate the unobserved flux.  The rest-frame spectral regions probed by these filters are not significantly affected by the line blanketing apparent in the blue end of the spectrum (see \S\ref{sec:spec} and Figure \ref{fig:BB}).  Therefore the blackbody parameters from fitting the flux from these filters reflect estimates of the underlying continuum.  Similarly, the inferred bolometric luminosity calculated here is an estimate of the total luminosity, using the blackbody fits to calculate a bolometric correction.  Given the strong suppression in the blue evident in the spectra, the implicit assumption is that this absorbed flux is re-emitted at longer wavelengths not probed by our observations.  We assume negligible internal host galaxy reddening, consistent with the results of our light curve modeling (see \S\ref{modeling}) and analysis of the host galaxy's properties (see \S\ref{sec:host}).  

The bolometric luminosity at the epochs of the single-band ground-based and {\it HST} observations was estimated by assuming constant colors measured from the most recent epoch with $gri$ data.  This is a reasonable assumption at phases $\gtrsim1-2$ months after peak when SLSNe exhibit little temperature evolution.  At earlier phases, however, rapid temperature evolution in SLSNe leads to rapidly varying colors.  We therefore consider our bolometric estimates for the first two epochs to be strict lower limits.  We find this is consistent with our magnetar modeling which predicts higher temperatures before and near peak than that suggested by our first multi-band observation.  For the same reason, the time of peak bolometric emission likely occurred before the peak in $w+r$ bands.  In Figure \ref{fig:bolLC} we show the temperature and radius evolution and corresponding bolometric light curve inferred from the model light curves shown in Figure \ref{fig:mosfit}.  Therefore due to our lack of multi-band data near peak and to facilitate comparisons, we use the results of our magnetar modeling to guide the phasing of the bolometric lightcurve.  

The bolometric light curve and photospheric temperature and radius evolution of SN\,2016inl are shown in Figure \ref{fig:bolLC}.  The bolometric light curve is presented relative to the time of explosion inferred from our magnetar modeling in \S\ref{modeling}, while the temperature and radius evolution are presented relative to the inferred time of bolometric maximum assuming the SED evolution inferred from our modeling.  By the time of our first multi-band observation, at $\approx28$ days after peak in $w+r$, SN\,2016inl exhibited a photospheric temperature of $\approx 9500$ K, followed by a drop to $\approx 8000$ K.  Subsequently, the temperature remains roughly constant, possibly increasing slightly, for about 200 days post-peak, and finally cools to about 7500 K in our final multi-band observations, consistent with the slow spectroscopic evolution.  Although the uncertainties are large, the possible late-time drop in temperature corresponds to a short phase in the light curve at $\sim 260$ rest-frame days when SN\,2016inl drops in luminosity relatively rapidly compared to the overall decline.  This is apparent in the multi-band data shown in Figure \ref{fig:obsLC} and the bolometric light curve.  The inferred photospheric radius recedes from about $5\times 10^{15}$ cm  to $\approx 10^{15}$ cm by about 200 rest-frame days.  We note that this temperature and radius evolution is similar to SN\,2015bn but different from PS1-14bj, which exhibited an increasing temperature.  

We find that the bolometric light curve of SN\,2016inl reached a peak of at least $\approx 1.3 \times 10^{44}$ erg s$^{-1}$, about 3 times brighter than PS1-14bj (Figure \ref{fig:bolLC}).  The inferred peak luminosity from our magnetar model is $\approx 2 \times 10^{44}$ erg s$^{-1}$.  The light curve is comparable in brightness and evolution to SN\,2015bn, exhibiting the same late-time flattening (linear behavior in log-log space) to a decline rate much slower than $^{56}$Co decay.  In Figure \ref{fig:bolLC} we also show $L\propto t^{-2}$, the decline predicted by magnetar spin-down assuming the thermalization of a fixed fraction of the magnetic dipole radiation \citep{Woosley2010}.  As the optical depth decreases, an increasing amount of high energy radiation is able to escape, leading to a steeper predicted decline of $L\propto t^{-4}$, also shown in the figure (this assumes a $\gamma$-ray opacity that is constant in time; \citealt{Nicholllate15bn}).  

As with the $r$+F625W light curve, we fit the bolometric light curve at times $\gtrsim200$ rest-frame days since peak with the function $L \propto t^{-\alpha}$, relative to the time of explosion determined from our magnetar modeling in \S\ref{modeling}.  We find a best-fit power-law slope of $\alpha = 2.9\pm 0.4$, consistent with the $r$+F625W fit.  The slope is shallower than $\alpha \approx 3.8$ observed for SN\,2015bn \citep{Nicholllate15bn}.

\section{Host Galaxy}
\label{sec:host}

As discussed in \S\ref{sec:LC} and as shown in Figure~\ref{fig:HST}, our late-time {\it HST} observations reveal a resolved host galaxy underlying the position of SN\,2016inl.  From our two-component GALFIT modeling we find the best-fit Sersic profile has an index of $n=0.9\pm 0.1$ (i.e., an exponential profile) and an effective radius of $r_{e} = 4.6 \pm 0.8$ kpc.  The corresponding model brightness of the galaxy in the F625W filter is $m_{\rm F625W}=23.26 \pm 0.24$, corrected for Galactic extinction.

We also obtained deep ground-based templates in December 2019 and January 2020 that yielded detections of the host galaxy in the $griz$ filters.  These data were obtained more than a year after the last ground-based non-detection of the SN, indicating they do not contain flux from the SN. Moreover, these templates were taken after the last {\it HST} observation when the SN had already faded below 26 mag in the F625W filter.  From these ground-based templates we measure using {\tt SExtractor} \citep{BertinArnouts1996} host galaxy magnitudes of $m_{g} = 23.75 \pm 0.08$, $m_{r} = 23.05 \pm 0.04$, $m_{i} = 22.90 \pm 0.06$, and $m_{z} = 22.96 \pm 0.12$ (corrected for Galactic extinction).  Using filter transformations from \citet{Jordi2006} and converting to the rest-frame (without a full K-correction), we find this corresponds to an absolute $B$-band luminosity of $M_{B}\approx -16.6$ mag, similar to the host galaxies of previous SLSNe \citep{Lunnan2014,Leloudas2015,Perley2016,Schulze2018}.  Assuming an empirical luminosity-metallicity relationship for dwarf galaxies \citep{Lee2006}, this corresponds to a metallicity of $12 + {\rm log(O/H)}\approx 8.1$.  

We further analyze the host galaxy emission lines detected in the GMOS spectrum at $+263$ days after peak.  We measure line fluxes by fitting Gaussian profiles after subtracting local fits to the SN spectrum.  We find the flux ratio H$\alpha$/H$\beta$ $=2.7 \pm 0.8$ is consistent with that predicted for Case B recombination, indicating no significant host galaxy extinction; this is consistent with our light curve modeling (Table~\ref{tab:param}).  We estimate the star formation rate (SFR) using the H$\alpha$ luminosity calibration from \citet{Kennicutt1998}, finding ${\rm SFR}\approx 0.14$ M$_{\odot}$ yr$^{-1}$, consistent with the SLSN host galaxy population \citep{Lunnan2014,Perley2016}.

\section{Discussion} 
\label{sec:disc}

SN\,2016inl is a rare SLSN from a massive progenitor, which represents only the second known event with a late-time decline rate slower than $^{56}$Co decay and exhibits the best spectral match with PS1-14bj, an SLSN with a very broad light curve consistent with PISNe.  The slow decline conclusively rules out PISN models as a possible explanation for SN\,2016inl, and by association, indicates PS1-14bj was also not a PISN explosion.  Here we discuss the implications of SN\,2016inl's shallow power-law decline in the context of the magnetar model, possible alternative explanations, and the engine, ejecta, and spectral properties of SN\,2016inl in the context of the SLSN population.

\subsection{Implications of the Light Curve Flattening}

SN\,2016inl's late-time light curve flattening is consistent with the expectation from magnetar energy input that the light curve will eventually settle on to a power law.  As shown in Figures \ref{fig:obsLC} and \ref{fig:bolLC}, SN\,2016inl exhibits a flattening well below the rate of $^{56}$Co and consistent with the power-law $L \propto t^{-2.8}$, shallower than that discovered in late-time observations of SN\,2015bn \citep{Nicholllate15bn}.  A fit to data beyond 200 days after peak shows SN\,2015bn tracks a $L \propto t^{-3.8}$ power law.    

At $\gtrsim$year timescales, when the ejecta are becoming optically thin to $\gamma$-rays, the evolving degree of thermalization is a key factor impacting the light curve slope.  Assuming complete thermalization, or if a fixed fraction of the input energy is thermalized, the light curve would track the engine input luminosity, $L\propto t^{-2}$, whereas decreasing thermalization leads to steeper power laws.  In the case of a constant $\gamma$-ray opacity, the light curve would track $L\propto t^{-4}$.  In {\tt MOSFiT}, $\gamma$-ray leakage is parameterized using $\kappa_{\gamma}$, a constant $\gamma$-ray opacity, which for SN\,2016inl is $\kappa_{\gamma}\approx 0.11$ cm$^{2}$ g$^{-1}$.  This is an order of magnitude larger than in SN\,2015bn, delaying the escape of $\gamma$-rays, and hence consistent with the shallower power-law decline.  We note that while the assumption of a constant $\kappa_{\gamma}$ is likely an oversimplification, this provides an adequate fit to the data.

Recently, \citet{VurmMetzger2021} presented the first radiative transfer models of magnetar-powered SN light curves, including the numerous photon-matter interactions that are possible in the magnetar wind nebula and expanding ejecta.  They find that the thermalization opacity varies with time, and that the nebula magnetization strongly affects its evolution.  In particular, they find that significant $\gamma$-ray leakage leading to light curves steeper than the engine input luminosity is only possible for low values of nebula magnetization.  This implies that substantial magnetic dissipation in the nebula is required to explain events like SN\,2015bn and SN\,2016inl.  While a numerical model using the engine parameters relevant for SN\,2016inl has not  been calculated by \citet{VurmMetzger2021}, its shallower power-law decline likely indicates a higher degreee of magnetization than in SN\,2015bn.

Another consequence of a varying $\gamma$-ray opacity predicted by \citet{VurmMetzger2021} is that the power-law index may change with time.  In particular, their model that best matches SN\,2015bn predicts a transition to a shallower power law during a phase in which the opacity increases; this effect may be discernible in the final two epochs of SN\,2015bn, both of which exceed the best-fit {\tt MOSFiT} model \citep{Nicholllate15bn}.  SN\,2016inl may similarly be in the process of transitioning to an even slower decline, motivating future observations with {\it HST} and/or {\it JWST}.  

Whether or not the power-law behavior observed in SN\,2016inl and SN\,2015bn is a ubiquitous feature of SLSN light curves will be a powerful test of the magnetar model, highlighting the importance of late-time observations.  The diversity already evident from the first two SLSNe with such data indicates that late-time optical observations may shed direct light on the properties of the engine.

\subsubsection{Alternative Explanations}

While the discovery of a second SLSN that follows a power-law decline at late times strengthens the case for a magnetar engine, it is important to consider other possible explanations.   

One potential source of flattening is a light echo, where SN light emitted at early times is reflected by dust sheets into the line of sight, arriving at the observer at later phases.  Such flattening has been observed in SN\,2006gy \citep{Miller2010} and some Type Ia SNe (\citealt{Patat2005} and references therein).  However, like SN\,2015bn, SN\,2016inl does not show any sign of early SN features in its latest spectrum or any trend in its color evolution consistent with contamination from reflected light, though we do not have color information beyond $\sim300$ rest-frame days when the flattening becomes readily apparent.  In addition, late-time flattening due to light echos is generally expected to be even slower than the flattening observed in SN\,2016inl ($0.1-0.3$ mag yr$^{-1}$; \citealt{Patat2005}).  

Another alternative explanation is late-time CSM interaction.  However, there is no evidence at any point during SN\,2016inl's evolution of CSM interaction.  Some SLSNe have shown spectral signatures of CSM interaction \citep{Yan2015,Yan2017b}, namely the emergence of broad H$\alpha$ emission lines at $\sim100-200$ days after peak interpreted as the result of ejecta interaction with a neutral hydrogen shell located at distances of $\sim10^{16}$ cm.  SN\,2016inl, on the other hand, shows no sign of an emerging broad component in its spectra, including the latest available at $+263$ days after peak (Figure \ref{fig:Halpha}).  The events showing broad H$\alpha$ are also generally distinct from other SLSNe, with notably weaker [\ion{O}{1}]$\lambda$6300 lines \citep{Yan2017b}.  SN\,2016inl is instead remarkably similar to SN\,2015bn at late phases, an SN for which even later spectra and radio limits conclusively rule out late-time CSM interaction \citep{Nicholllate15bn}.  Therefore the lack of spectroscopic evidence for CSM interaction in SN\,2016inl coupled with its strong similarity to SN\,2015bn, makes CSM interaction unlikely as an explanation for the flattening in SN\,2016inl.

\subsection{The Diversity of SLSNe}

Motivated by the spectral similarity of SN\,2016inl and PS1-14bj, and by the late-time light curve similarity to SN\,2015bn, we compare here the engine and ejecta parameters of these three events (Figure~\ref{fig:param}) and compare their spectral evolution.

\begin{figure}
\centering
\includegraphics[scale=0.37]{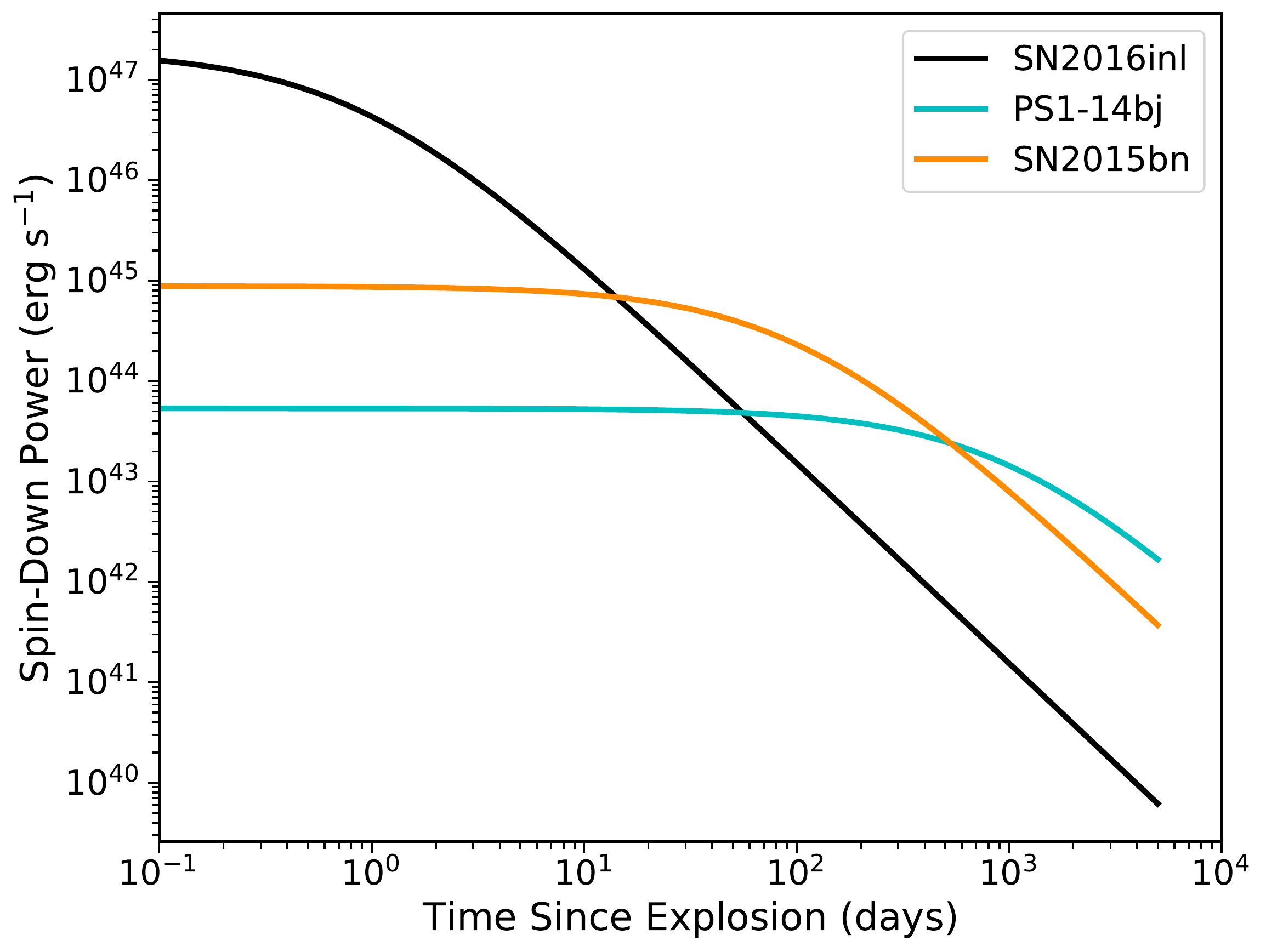}
\caption{Spin-down power corresponding to the inferred engine parameters for SN\,2016inl, PS1-14bj, and SN\,2015bn.  SN\,2016inl's engine spins down rapidly compared to PS1-14bj's, and consequently has a higher peak luminosity and predicted shorter rise time.  The model therefore predicts a higher temperature at peak, matching the temperature of PS1-14bj after having cooled, consistent with our earliest spectrum.}
\label{fig:spindown}
\end{figure}

SN\,2016inl's larger luminosity compared to PS1-14bj can be understood as being the result of a more powerful central engine; indeed, SN\,2016inl has a faster initial spin than PS1-14bj (1.5 ms versus 3 ms) giving it a reservoir of rotational energy $\approx 3$ times larger.  This, in conjunction, with a magnetic field that is larger by a factor of about 20, means that SN\,2016inl's engine spins down more rapidly than that powering PS1-14bj, with engine timescales of $\approx 1$ and $\approx10^3$ days, respectively.  In Figure \ref{fig:spindown} we show the spin-down power for SN\,2016inl and PS1-14bj.  Although SN\,2016inl's engine spins down rapidly compared to its diffusion time of $\approx80$ days, the input luminosity is still quite high ($\approx10^{44}$ erg s$^{-1}$) by the time SN\,2016inl reaches maximum brightness.  With a similar diffusion time but much longer spin-down time, PS1-14bj exhibits a longer rise ($\approx125$ days) but lower peak luminosity.  We also show SN\,2015bn in Figure \ref{fig:spindown} which has an intermediate engine timescale of $\approx10^2$ days, well matched to its diffusion time of $\approx50$ days, enabling it to reach the highest luminosity of the three events. 

The temperature evolution inferred from our spectra is consistent with this model.  Our first spectrum of SN\,2016inl was taken at $+12$ days after $r$-band maximum, when the photospheric temperature was $\approx10000$ K.  Although we have no temperature information prior to this date, our light curve model predicts a higher temperature of $\approx15000$ K at bolometric maximum, which occurred $\approx40$ days before our first spectrum (see Figure \ref{fig:bolLC}).  This is hotter than PS1-14bj at bolometric peak, consistent with SN\,2016inl's larger luminosity and shorter rise time.  Subsequently, after having cooled slightly, SN\,2016inl's spectrum taken after bolometric peak matches the temperature of PS1-14bj.  SN\,2016inl's temperature evolution is instead more comparable to SN\,2015bn.  These comparisons further highlight the importance of temperature (itself impacted by the location of the photosphere and luminosity) rather than phase for understanding the spectra of SLSNe, as discussed in \citet{Nicholl2017}.  

Importantly, PS1-14bj and SN\,2016inl, are easily understood in the context of the magnetar model, and do not show evidence for a separate explosion channel or power source than other SLSNe.  Their broad light curves are simply an indication of a higher than average ejecta mass, which coupled with variations in engine timescale (itself controlled by two variables), lead to variations in rise time, peak luminosity, and temperature evolution.     

In detail, there are other factors affecting the spectral characteristics of SLSNe, notably the abundance of Fe-group elements.  This additional variable can make even rapidly evolving, low ejecta mass events, which have hot compact photospheres at peak, appear redder due to Fe-group opacity.  In \citet{Blanchard2019}, we found significant absorption from \ion{Co}{2} in SN\,2017dwh near peak, followed by a rapid transition to the red spectrum typical of broad-lined Type Ic SNe after peak.  Both SN\,2016inl and PS1-14bj also show evidence for enhanced Fe-group absorption that makes the overall shape of their early spectra markedly different from other events (Figures \ref{fig:spec} and \ref{fig:BB}).  While the cause of the varying amount of Fe-group absorption in SLSNe is not known, \citet{Blanchard2019} suggest that it may be caused by enhanced mixing due to a jet-like outflow launched by the central engine.  If so, SN\,2016inl and PS1-14bj indicate such jets may be possible from engines with a wide range of magnetic field strengths. 

Finally, at late times ($\gtrsim 250$ days), when the photosphere has receded and the ejecta start to become optically thin, SN\,2016inl, PS1-14bj and SN\,2015bn become more spectroscopically similar.  This indicates that they share similar ejecta properties despite the impact of different engine parameters at early time.  Differences in some nebular emission line strengths likely reflect the wide range of possible progenitor masses \citep{Blanchard2020}. For example, the smaller ratio of [\ion{Ca}{2}]$\lambda7300$/[\ion{O}{1}]$\lambda6300$ in SN\,2016inl than that in SN\,2015bn may be indicative of a larger core mass, consistent with the higher ejecta mass inferred from the light curve modeling.

\section{Summary and Conclusions}
\label{sec:conc}

We presented optical observations, spanning to about 700 rest-frame days, of SN\,2016inl, a slowly-evolving hydrogen-poor SLSN at $z=0.3057$.  Our key findings are: 

\begin{itemize}
    \item SN\,2016inl closely matches the red spectra of PS1-14bj, with both events exhibiting a steep drop in flux below $\approx4000$ \AA, likely due to enhanced Fe-group absorption.
    
    \item SN\,2016inl is only the second SLSN (along with SN\,2015bn) with sufficient late-time data to exhibit a significant flattening in the decline rate, eventually reaching a rate much slower than the radioactive decay rate of $^{56}$Co.
    
    \item From fitting the late-time $r$-band light curve, we find that the decline is described by a power law, $L \propto t^{-2.8\pm0.2}$, shallower than the $L\propto t^{-3.8}$ decline found for SN\,2015bn.  
    
    \item Using a magnetar engine model we find that SN\,2016inl has engine parameters that are consistent with the SLSN population, but with a larger than average ejecta mass of $\approx 20$ M$_\odot$.
\end{itemize}

Based on the late-time light curve flattening we rule out a PISN model for SN\,2016inl, which remained viable for PS1-14bj due to the lack of data beyond 400 rest-frame days.  Previously, observations of the rising phase of the light curve have been used to argue against radioactive decay as the dominant energy source in slowly evolving SLSNe \citep{Nicholl2013}.  Our observations of SN\,2016inl indicate that even for events that lack detailed measurements on the rise, late-time observations can be used similarly to rule out radioactive decay.

A power-law decline is expected in the magnetar model, though the precise slope and its evolution are sensitive to the evolving degree of thermalization, controlled by the $\gamma$-ray opacity.  Although in {\tt MOSFiT} we assume a constant opacity, the higher value we find for SN\,2016inl compared to SN\,2015bn is consistent with the former's shallower slope.  In comparison with recent radiative transfer models \citep{VurmMetzger2021}, SN\,2016inl's shallower decline may be indicative of a higher nebula magnetization than in SN\,2015bn.  

The evidence for a magnetar engine in SN\,2016inl suggests by association that PS1-14bj was also powered by a central engine.  Indeed, both events fit within the trends seen in the full population.  Their large ejecta masses are responsible for their broad light curves, with a variable engine timescale leading to variation in the peak luminosity and temperature evolution.  Moreover, both events show enhanced Fe-group absorption compared to the general SLSN population, highlighting the need for theoretical models that can explain the formation and mixing of such elements.  Importantly, SN\,2016inl provides a link between SLSNe with diverse behavior (e.g., PS1-14bj, SN\,2015bn, SN\,2017dwh), and further supports the idea that the significant diversity exhibited by SLSNe is a natural consequence of the magnetar model.

\acknowledgments
P.K.B.~is supported by a CIERA Postdoctoral Fellowship.  The Berger Time-Domain Group at Harvard is supported by NSF and NASA grants. M.N.~is supported by a Royal Astronomical Society Research Fellowship and by funding from the European Research Council (ERC) under the European Union’s Horizon 2020 research and innovation programme (grant agreement No.~948381). This research is based in part on observations made with the NASA/ESA {\it Hubble Space Telescope} obtained from the Space Telescope Science Institute, which is operated by the Association of Universities for Research in Astronomy, Inc., under NASA contract NAS 5–26555. These observations are associated with programs 15162 and 15496.  This work is based in part on observations obtained at the MDM Observatory, operated by Dartmouth College, Columbia University, Ohio State University, Ohio University, and the University of Michigan. We thank J. Klusmeyer for acquiring the MDM queue observations.  This paper uses data products produced by the OIR Telescope Data Center, supported by the Smithsonian Astrophysical Observatory.  Some observations reported here were obtained at the MMT Observatory, a joint facility of the Smithsonian Institution and the University of Arizona.  This paper includes data gathered with the 6.5 meter Magellan Telescopes located at Las Campanas Observatory, Chile.  Based in part on observations obtained at the international Gemini Observatory, a program of NSF’s NOIRLab, which is managed by the Association of Universities for Research in Astronomy (AURA) under a cooperative agreement with the National Science Foundation on behalf of the Gemini Observatory partnership: the National Science Foundation (United States), National Research Council (Canada), Agencia Nacional de Investigaci\'{o}n y Desarrollo (Chile), Ministerio de Ciencia, Tecnolog\'{i}a e Innovaci\'{o}n (Argentina), Minist\'{e}rio da Ci\^{e}ncia, Tecnologia, Inova\c{c}\~{o}es e Comunica\c{c}\~{o}es (Brazil), and Korea Astronomy and Space Science Institute (Republic of Korea).  These observations are associated with Gemini program GN-2017B-FT-5.

\end{document}